\documentclass[]{article}
\usepackage{graphicx}
\makeindex
\begin{document}

\baselineskip 2.5ex
\parskip 3.5ex
\newcommand{\noi}{\noindent}
\renewcommand{\thesection}{\arabic{section}}
\renewcommand{\thesubsection}{\arabic{section}.\Alph{subsection}}
\renewcommand{\theequation}{\arabic{section}.\arabic{equation}}
\renewcommand{\a}{\alpha}
\renewcommand{\b}{\beta}
\newcommand{\e}{\epsilon}
\newcommand{\g}{\gamma}
\newcommand{\G}{\Gamma}
\renewcommand{\d}{\delta}
\newcommand{\D}{\Delta}
\renewcommand{\vec}{\bm}
\newcommand{\svec}[1]{\mbox{{\footnotesize $\bm{#1}$}}}
\newcommand{\bm}[1]{\mbox{\boldmath $#1$}}
\newcommand{\sumc}[3]{\sum_{#1=#2}^{#3}}
\newcommand{\pder}[2]{\frac{\partial {#1}}{\partial {#2}}}
\newcommand{\pdert}[2]{\frac{\partial^2 {#1}}{\partial {#2}^2}}
\newcommand{\fder}[2]{\frac{\delta {#1}}{\delta {#2}}}
\newcommand{\PDD}[3]{\left.\frac{\partial^{2}{#1}}{\partial{#2}^{2}}\right|_
{#3}}
\newcommand{\PD}[3]{\left.\frac{\partial{#1}}{\partial{#2}}\right|_{#3}}
\newcommand{\der}[2]{\frac{d {#1}}{d {#2}}}
\newcommand{\diff}[1]{{\rm d}{#1}}
\newcommand{\fif}{\mbox{$\Longleftrightarrow$}}
\newcommand{\hsk}{\hspace{2em}}
\newcommand{\beq}{\begin{equation}}
\newcommand{\bequo}{\begin{quotation}}
\newcommand{\beqa}{\begin{eqnarray}}
\newcommand{\eeq}{\end{equation}}
\newcommand{\equo}{\end{quotation}}
\newcommand{\eeqa}{\end{eqnarray}}
\newcommand{\la}{\langle}
\newcommand{\ra}{\rangle}
\newcommand{\hsp}{\hspace*{8em}}
\newcommand{\rate}[2]{I_{\tilde{#1}}^{(#2)}}
\newcommand{\lima}[1]{\lim_{#1 \rightarrow \infty} \frac{1}{#1} \ln}
\newcommand{\meas}[2]{{#1}(\mbox{d} {#2})}
\newcommand{\limin}[1]{\lim_{#1 \rightarrow \infty}}
\newcommand{\un}{\underline}
\newcommand{\dd}{\mbox{d}}
\newcommand{\lb}{\label}
\newcommand{\fr}[1]{(\ref{#1})}
\newcommand{\non}{\nonumber}
\newcommand{\bmc}[1]{\mbox{\boldmath $#1$}^{C}}
\newcommand{\bml}[1]{\mbox{\boldmath $#1$}^{L}}
\newcommand{\sbm}[1]{\small{\mbox{\boldmath $#1$}}}
\newcommand{\fns}{\footnotesize}
\newcommand{\cvec}[3]{\left( \begin{array}{c} #1\\#2\\#3 \end{array} \right)}
\newcommand{\matr}[9]{\left(\begin{array}{ccc} #1&#2&#3\\#4&#5&#6\\#7&#8&#9
\end{array} \right)}
\newcommand{\see}{$\rightarrow$}
\newcommand{\map}{\rightarrow}
\newcommand{\maps}{\rightarrow}
\newcommand{\imply}{\Rightarrow}
\newcommand{\implies}{\Rightarrow}
\newcommand{\define}{\,\raisebox{-.4ex}{$\leftharpoondown$}{\hspace{-1.0em}\
raisebox{.45ex}{=\hspace{-.6em}=}\,}}
\newcommand{\defines}{\,\raisebox{-.4ex}{$\leftharpoondown$}{\hspace{-1.0em}
\raisebox{.45ex}{=\hspace{-.6em}=}\,}}
\newcommand{\bra}[1]{\langle{#1}|}
\newcommand{\ket}[1]{|{#1}\rangle}
\renewcommand{\i}[1]{\index{#1}}

\newcommand{\be}{\begin{equation}}
\newcommand{\benn}{\begin{eqnarray*}}
\newcommand{\eenn}{\end{eqnarray*}}
\newcommand{\ba}{\begin{array}}
\newcommand{\bea}{\begin{eqnarray}}
\newcommand{\ee}{\end{equation}}
\newcommand{\ea}{\end{array}}
\newcommand{\eea}{\end{eqnarray}}
\def\real{\hbox{\rm\setbox1=\hbox{I}\copy1\kern-.45\wd1 R}}
\def\nonn{\hbox{\rm\setbox1=\hbox{I}\copy1\kern-.45\wd1 N}}
\def\forc{\hbox{\rm\setbox1=\hbox{I}\copy1\kern-.45\wd1 F}}
\def\flux{\hbox{\setbox1=\hbox{f}\copy1\kern-.45\wd1 f}}
\newcommand{\al}{\alpha}
\newcommand{\bet}{\beta}
\newcommand{\gam}{\gamma}
\newcommand{\lam}{\lambda}
\newcommand{\eps}{\epsilon}
\newcommand{\ov}{\overline}
\newcommand{\Lrar}{\Leftrightarrow}
\newcommand{\rar}{\rightarrow}

\newtheorem{entry}{}[section]
\newcommand{\bent}[1]{\vspace*{-2cm}\hspace*{-1cm}\begin{entry}\lb{e{#1}}\rm}
\newcommand{\eent}{\end{entry}}
\newcommand{\fre}[1]{{\bf\ref{e{#1}}}}

\begin{center}

{\Large 
Temporal patterns of gene expression via nonmetric multidimensional scaling analysis}
\end{center}

running head: Temporal patterns via nonmetric MDS

Y.-h. Taguchi$^{1,2,*}$ and  Y. Oono$^3$,

\noi
$^1$Department of Physics, Faculty of Science and Technology,
Chuo University, 1-13-27 Kasuga, Bunkyo-ku, Tokyo 112-8551, Japan,\\
$^2$Institute for Science and Technology,
Chuo University, 1-13-27 Kasuga, Bunkyo-ku, Tokyo 112-8551, Japan,\\
and\\
$^3$Department of Physics,  1110 W. Green Street,
       University of Illinois at Urbana-Champaign,
       Urbana, IL 61801-3080, USA.

\medskip

\hrule
\noi
* To whom correspondence should be addressed

\pagebreak

\twocolumn
\noi
{\bf ABSTRACT}\\
{\bf Motivation}: Microarray experiments result in large scale 
data sets that require extensive mining and refining to extract 
useful information.  We have been developing an
efficient novel 
algorithm for nonmetric multidimensional scaling (nMDS) analysis
for very large data sets as a maximally unsupervised data mining
device.  We wish to demonstrate its usefulness in the context of
bioinformatics. In our motivation is also an aim to demonstrate
that intrinsically nonlinear methods are generally advantageous
in data mining.\\
{\bf Results}: The Pearson correlation distance measure is used
to indicate the dissimilarity of the gene activities in 
transcriptional response of cell cycle-synchronized human 
fibroblasts to serum [Iyer {\it et al}., Science {\bf 283}, 83 (1999)]. 
These dissimilarity data have been analyzed with our nMDS 
algorithm to produce an almost circular arrangement of the genes.
The temporal expression patterns of the genes rotate along this
circular arrangement.  If an appropriate preparation procedure 
may be applied to the original data set, linear methods such as the
principal component analysis (PCA) could achieve reasonable results, 
but without data preprocessing linear methods such as PCA cannot 
achieve a useful picture.  Furthermore, even with an appropriate data 
preprocessing, the outcomes of linear procedures are not as clearcut 
as those by nMDS without preprocessing. 
\\
\noi
{\bf Availability:}
The fortran source code of the method used in this analysis  (`pure 
nMDS')  is available at \\  http://www.granular.com/MDS/\\
\noi
{\bf Contact:}
tag@granular.com ; yoono@uiuc.edu\\

\noi
{\bf INTRODUCTION}\\
Each DNA microarray experiment can give us information about 
the relative populations of mRNAs for thousands of genes.   This
implies  that without extensive data mining it is often hard to 
recognize any useful information  from the experimental results.
In this paper we demonstrate that a nonmetric multidimensional
scaling (nMDS) method can be a powerful unsupervised means to
extract temporal expression patterns of genes.  A data mining 
procedure may be useful, if it is flexible enough to incorporate 
any level of supervision, but we believe that the most basic 
feature required for any good data mining method is to be able to
extract recognizable patterns reproducibly without supervision.
In this sense our nMDS method is clearly demonstrated to be a 
useful means of data mining.

We have been developing an efficient nMDS 
technique for large data sets (Taguchi and Oono, 1999, Taguchi {\it et al}., 2001).  The 
input is the rank order of (dis)similarities among the objects (in
the present case, genes). Our algorithm is  maximally nonmetric
in the sense that any introduction of intermediate metric 
coordinates obtained by monotone regression common to the 
conventional nMDS methods is avoided.

Data compression is essentially a problem of linear functional 
analysis as Donoho {\it et al}.\ (1998) stresses.  In contrast, we
believe data mining is essentially nonlinear.  There are linear 
algebraic methods such as the principal component analysis (PCA)
for data mining, but it is expected that nonlinear methods are, in
principle, more powerful.  The present paper illustrates this 
point.  Indeed, in our case PCA cannot find any comprehensible 
temporal pattern in low dimensional spaces without an 
appropriate data preparation.\\

\noi
{\bf SYSTEMS AND METHODS}\\
\noi
{\bf Systems to Analyze}\\
The gene activities in transcriptional response of cell 
cycle-synchronized human fibroblasts to serum reported by Iyer
{\it et al}.\ (1999) are analyzed.  The microarray data used in this 
analysis is available at http://genome-www.stanford.edu/serum/.\\

\noi
{\bf Other Possible Analysis Methods}\\
To extract interpretable patterns from microarray  data, cluster
and linear 
multivariate analyses seem to be the two major strategies. 
However, these methods may not be ideally suited for the purpose.

The cluster analysis seems to be the most popular analytical 
method  (Slonim, 2002).
For example, the hierarchical clustering method seems to be popular
(Eisen {\it et al}., 1998).
Perhaps there are two fundamental criticisms against clustering methods. 
Classifying the expression patterns as functions of time is often attempted 
by clustering methods (Spellman {\it et al} 1998). However,  it is often the 
case that temporal gene expression patterns vary rather 
continuously without natural gaps among various patterns; needless to say, 
cluster analysis is not a suitable method to classify continuously changing 
objects.  The second criticism is that clustering methods cannot give any 
relation among resultant clusters other than `genealogical relations' mimicking 
similarities. Therefore, clustering is unsuitable for temporal pattern analysis.  
For example, if the resulting clustering is ((A,B),C), it is only by inspection that 
ABC or BAC is chosen as a natural temporal pattern or structure.  Thus, to exhibit the 
temporal pattern
rearranging the genes in each cluster by hand is needed.  An 
example with such a procedure may be found in Spellman {\it et al}.\ (1998).  

Perhaps the most popular linear multivariate analysis method
is the principal component analysis (PCA). The main idea is to choose a
data-adapted basis set, and to make a subspace that can capture
salient features of the original data set.  In principle, the method could
capture  the temporal order in the gene expression pattern, but 
the dimension of the subspace may not be low even if the 
data are on a very low dimensional manifold.  In short, the information compression capability of linear methods is 
generally feeble.  This can be well illustrated by the data we
wish to analyze in this paper: PCA cannot capture any clear 
temporal order as shown in Fig.\ 1, where the two dimensional space 
spanned by the first two principal components is shown there.  
The temporal expression pattern is hardly seen from the result.   

\marginparsep -9mm
\marginpar{\hspace{10cm}\fbox{Fig. 1}}
\marginparsep 4mm

However, apparently, Holter {\it et al}.\ (2000) demonstrated that the singular value 
decomposition (SVD; a linear method) is remarkably successful in 
extracting the characteristic modes.   The reader must wonder
why there is a difference between this result and the one due to
PCA that is not successful.  The secret is in the highly nonlinear
`polishing' of the original data (proposed by Eisen {\it et al}. (1998)).  
However, the role of this  nonlinear polishing must be considered
carefully, because it can generate a spurious temporal behavior.
Therefore, we relegate the comparison of these linear methods 
with data preprocessing and our nMDS to Appendix II.  The salient
conclusions are:\\
(1) Linear methods such as
PCA and SVD could perhaps achieve reasonable results,
if a data preparation scheme is chosen correctly.
However, best linear results are generally fairly inferior to nonlinear
results.\\
(2) The data preparation such as the `polishing' used by Holter {\it et
al.}\  (2000) could actually corrupt the original data (as illustrated in
Appendix II), and should be avoided.

An interesting proposal is to use the partial least squares (PLS)
regression (Johansson {\it et al}, 2003). In this case one may 
assume a temporal order one wishes to extract (say, a sinusoidal
change in time), and the original data are organized around the 
expected pattern.  This is, so to speak, to analyze the data 
according to a certain prejudice.  Although in the process of 
organization no supervision is needed, the pattern to be extracted
(that is, the `prejudice') must be presupposed.  Thus, even if it is
unsupervised, it is hardly a foolproof method.  Furthermore, if a
clear objective pattern could be extractable by this method, 
certainly nMDS can achieve the same goal without any 
presupposed pattern required by PLS.

Metric multidimensional scaling methods (MDS) may also be used, but 
it depends on the definition of the dissimilarity. Therefore, unless the 
measure of dissimilarity is (almost) dictated by the data or by the 
context of the data analysis, arbitrary elements are introduced. For 
example, in the case of the microarray data there is no natural 
dissimilarity measure, so the metric that may capture detailed 
information could carry spurious information (disinformation, so to 
speak) as well.  Also, if the dissimilarity data is with signs as in the 
case of correlation coefficients, an extra arbitrary factor intervenes 
when they are converted to positive dissimilarity measures.  A further 
disadvantage of metric MDS (and of linear methods) is that these methods are 
vulnerable to missing or grossly inaccurate data.

The cluster analysis with the aid of self-organizing maps (SOM) is definitely 
a nonlinear data analysis method, but as we have seen in Kasturi {\it et al}.\ 
(2003) it is not particularly suitable for extracting temporal order.  Kasturi 
{\it et al}.\ comment that SOM is not particularly better than the ordinary 
cluster analysis.  Furthermore, as can be seen from the fact that the use of 
a particular initial condition can be a methodological paper (Kanaya {\em et al}., 
2001), we must worry about the {\em ad hoc} initial-condition dependence of 
the results. 
\\

\noi
{\bf ALGORITHM}\\
\noi
{\bf Basic idea of algorithm for nMDS}\\
The philosophy of nMDS  (Shepard 1962a, 1962b, Kruskal 1964a, 1964b)
is to find a
constellation in a certain space ${\cal R}$ of points representing
the objects under study (genes in the present case) such that the
pairwise distances $d$ of  the points in ${\cal R}$ have the rank
order in closest agreement with the rank order of the pairwise 
dissimilarities $\d$  of the corresponding objects that are given
as the raw (or the original)  input data.   

The conventional nMDS methods assume a certain intermediate 
pair distance  $\hat{d}$ that is chosen  as close as possible to $d$
for a given object pair under the condition that it is monotone 
with respect to the given actual ordering of the dissimilarities
$\d$.  The choice of $\hat{d}$ is not unique.  The discrepancy 
between $d$ and $\hat{d}$ is called the stress, and all the 
algorithms attempt to minimize it. Depending on the choice of $\hat{d}$ and on the interpretation of
``as close as,'' different methods have been 
proposed (see, for example, Green {\it et al}. (1970), Cox and Cox
(1994), and Borg and Groenen (1997) ). 
The choice of $\hat{d}$ affects the outcome.  $\hat{d}$ is required
only by technical reasons for implementation
of the basic nonmetric idea, so to be faithful to
the original idea due to Shepard (1962a, 1962b) we must compare $\d$ with $d$
directly.  Our motivation is to make an algorithm that is 
maximally nonmetric in the sense that we get rid of $\hat{d}$.

The basic idea of this `purely nonmetric' algorithm is as 
follows (Taguchi and Oono, 1999, Taguchi {\it et al}., 2001):
in a metric space ${\cal R}$ (in this paper, 
$D$-dimensional Euclidean space ${\bm R}^D$ is used) $N$ points
representing the $N$ objects 
are placed as an initial configuration. 
For this initial trial configuration we compute the pair distances
$d(i,j)$, and then rank them according to their magnitudes. 
Comparing this ranking and that according to the dissimilarity 
$\delta(i,j)$, we compute the `force' that moves the points in 
${\cal R}$ to reduce the discrepancies between these two 
rankings.  After moving the points according to the `forces', the
new `forces' are computed again, and the whole adjusting process
of the object positions in ${\cal R}$ is iterated until they 
converge sufficiently.  The details are in Appendix I.  

nMDS can usually recover 
geometrical objects correctly (up to scaling, orientation, and direction) when there are
sufficiently many (say, $\ge 30$) objects. Therefore, nMDS is a 
versatile multivariate analysis method. 

It is desirable to have a criterion for convergence (analogous to
the level of the stress in the conventional nMDS), or a measure of
goodness of embedding.  To this end let us recall the Kendall 
statistics $K$ (p364, Hollander and Wolfe (1999)),
\[
K = \sum_{\la\a,\b\ra} \mbox{sign}[(d_\a-d_\b)(\d_\a-\d_\b)],
\]
where the summation is over all the pairs of dissimilarities (distances between objects) $\la \a, \b\ra$ (i.e., $\a$ (also $\b$) denotes  a pair of objects). 
Usually, this is 
used for a statistical test to reject the null hypothesis that 
$\{d(i,j)\}$ does not correlate with $\{\d(i,j)\}$ (The 
contribution of ties is negligible usually for large data set, so we
do not pay any particular attention to tie data).

Here, we use this value to estimate the number of the objects 
embedded correctly. 
If all the
objects are correctly embedded, all the summands are $1$. Thus,
if $N'$ objects are correctly embedded, and if we may assume that
the rest are uncorrelated, then $K$ is expected to be 
\[
K > n'(n'-1)/2 - O[N^2],
\]
where $n' = N'(N'-1)/2$, and the subtraction comes from the 
random sum of at most $_{_{N}C_{2}}C_2$ of $\pm1$.  If the embedding is
successful for the majority of the objects, then $n' = O[N^2]$, so
we may ignore the contribution of the bad points. Thus, we may estimate
\[
N' \simeq \sqrt{2\sqrt{2K}}.
\]
Therefore, we adopt $100\sqrt{2\sqrt{2K}}/N$\%  as an indicator of
goodness of embedding.

We must also discuss the initial configuration dependence of the
result. Our algorithm is not free from the problem of local minima
as all of the previously proposed algorithms for nMDS and as high
dimensional nonlinear optimization problems in general.  
However, generally speaking, this 
dependence has only a very minor effect. 
This will be checked for the fibroblast data (See below).

\noi
{\bf RESULTS}\\
We have found that the fibroblast data may be embedded in a two
dimensional space roughly as a ring (Fig.\ 2).  The estimated number
of correctly embedded genes is about 480 among all the 517 genes
(i.e., the goodness of embedding is more than 90\%). (Also 516 out
of 517 genes have $P < 0.005$ confidence level (see
Appendix I) ). Thus, we  conclude that the obtained configuration is
sufficiently reliable.

\marginpar{\fbox{Fig. 2}}

\marginpar{\fbox{Fig. 3}}

This is in remarkable contrast with the PCA result mentioned 
already (Fig.\ 1).  Further remarkable is the fact that this ring-like 
arrangement of the genes faithfully represents the temporal 
expression patterns of the genes as can be seen clearly from the
rotation of the expression peaks around the ring (Fig.\ 3a).  It is 
noteworthy that the angle coordinate assigned to the genes 
according to the result shown in Fig.\ 2 automatically gives the
figure usually obtained through detailed Fourier analysis  (Fig.\
3b). These figures should be eloquent enough to attest to the 
usefulness of nMDS, a nonlinear data mining method.

Finally, to see the initial configuration dependence for the case of  the 
fibroblast data we constructed two 2D embedding results starting from 
two different random initial configurations. With the aid of the Procrustean 
similarity transformation (Borg and Groenen, 1997) one result is fit to the 
other (notice that our procedure is nonmetric, so to compare two independent 
results, appropriate scales, orientations, etc., must be optimally chosen).  
Fig.\ 4 demonstrates the close agreements of $x$- and $y$-coordinates of 
the two results.  As illustrated, the dependence on the initial conditions is 
very weak, and we may regard the embedded structure as a faithful 
representation of the information in the original data.

\marginparsep -9mm
\marginpar{\hspace{10cm}\fbox{Fig. 4}}
\marginparsep 4mm

As has been clearly demonstrated,  the 2D embedding is 
statistically natural and informative.  Still the 2D embedding is
not perfect, so it is interesting to see what we might obtain by
`unfolding' the 2D data, adding one more axis.   The unfolded result
is shown in Fig.\ 5. Here, the angular coordinates $\phi$ and 
$\theta$ of the spherical coordinate system is determined by the
$xy$-plane whose $x$-(resp., $y$-)axis is the first (resp., the 
second) principal component of the 3D embedded result.  The total
contribution of these two components is 86\%.  We do not 
recognize any clear pattern other than that captured in the 2D 
space.  Therefore, we may conclude that the 2D embedding result
is sufficiently reliable and informative.

\marginparsep -9mm
\marginpar{\fbox{Fig. 5}}
\marginparsep 4mm

\noi{\bf CONCLUSION}\\
We have demonstrated that the nMDS can be a useful tool for data
mining.  It is unsupervised, and perhaps maximally nonlinear.  Our
algorithm is probably the simplest among the nonmetric MDS 
algorithms and is efficient enough to enable the analysis of a few thousand objects
with a desktop PC.     

The NMDS algorithm works on the binary
relations among the objects, so if there are $N$ objects, 
computational complexity is of order $N^2$ at least.  Therefore, it is far
slower than linear methods such as PCA, although our nonlinear
algorithm is practically fast enough, 
because we have used for this work a small
notebook PC (Mobile Celeron 650MHz cpu with 256MB RAM).  
As pointed out and as has been 
illustrated, with an appropriate data preprocessing a certain linear
method could give us a reasonable result with less computational
efforts.  Although in this paper we have not made any particular effort to
reduce computational requirements, a practical way to use nMDS 
may be to prepare an initial configuration by a linear method with
an appropriate data preprocessing method that is verified to be 
consistent with the full nMDS results.

\noi
{\bf APPENDIX I}\\
\noi
{\bf `Purely' non-metric MDS algorithm}\\
Suppose $d(i,j)$ is the distance between objects $i$ and $j$ in 
${\cal R}$.  Let the ranking of $\d(i,j)$ among all the input 
dissimilarity data be $n$ and that of $d(i,j)$ among all the distances
between embedded pairs be $T_n$. If $n>T_n$ (resp., $n < T_n$),
we wish to `push' the pair $i$ and $j$ farther apart (resp., closer) in ${\cal R}$.
Intuitively speaking, to this end we introduce an `overdamped 
dynamics' of the points in ${\cal R}$  driven by the following 
potential function
\[
\Delta \equiv \sum(T_n -n)^2. \label{Dd}
\]
Here, the summation is over all the pairs (in the actual 
implementation of the algorithm, simpler forces are adopted than
the one obtained from this potential as seen below). This $\Delta$ may be 
regarded as a counterpart of the stress in the conventional nMDS.
As we will see later we can use quantities related to $\Delta$ to
evaluate the confidence level of the 
resultant configuration. Thus, an important feature of our nMDS
algorithm is that the optimization process is directly connected
to a process that improves the confidence level of the resultant
configuration.

The `pure nMDS' algorithm for $N$ objects 
may be described as follows:
\begin{enumerate}
\item Dissimilarities $\delta_{ij}$ $(i,j = 1, \cdots, N)$
   for $N$ objects are given. Order them as follows:
\[
   \cdots \leq \delta_{ij} \leq \delta_{kl} \leq \cdots.\label{delta} 
\]
\item Put $N$ points randomly in ${\cal R}$ as an
   initial configuration. 
\item Scale the position vectors in ${\cal R}$ such as 
$\sqrt{\sum_i {|\vec r}_i|^2}=1$,    where ${\vec r}_i$ is the 
current position of object $i$ in ${\cal R}$.
\item Compute $d_{ij}$ for all object pairs $(i,j)$
   in ${\cal R}$, and then order them as
\[
   \cdots \leq d_{ij} \leq d_{kl} \leq \cdots.\label{do}
\]
\item Suppose $\delta_{ij}$ is the $m$th largest in
   the ordering in \ref{delta} and $d_{ij}$ is the $T_m$th
   largest in the ordering  in \ref{do}. Assign $C_{ij}= T_m-m$. 
Calculate the following displacement 
   vector for $i$:
\[
   \d\vec{r}_i = s \sum_j C_{ij} \frac{{\vec r}_i-{\vec r}_j}{|{\vec r}_i-{\vec r}_j|},
\]
where $s=0.1 \times N^{-3}$ typically, and update $\vec{r}_i \map \vec{r}_i + \d \vec{r}_i$. 
\item Return to 3, and continue until the ``potential energy'' becomes
sufficiently small. 
\end{enumerate}

The reader may worry about the handling of  tie data.  
Generally speaking, for a large data set the fraction of tie 
relations is not significant; furthermore, if the result depends on
the handling schemes of tie data, the result is unreliable anyway.
Therefore, we do not pay any particular attention to the tie data
problem.

In the above algorithm, $s$ is a constant value.  In practice, we
could choose an appropriate schedule to vary $s$ as is often done
in optimization processes.  In this paper, for simplicity, we do not
attempt such a fine tuning.

In the above algorithm, we can deal with asymmetric data as 
well, i.e., $\delta_{ij} \neq \delta_{ji}$ if we compare 
$\delta_{ij}$ with $d_{ij}$ while $\delta_{ji}$ with $d_{ji} 
(=d_{ij})$.  Needless to say, if the mismatch between 
$\delta_{ij}$ and $\delta_{ji}$  is large, then representing the 
pair by a pair of points in a metric space is questionable.  
Therefore, we will not discuss this problem any further in this paper.

\noi
{\bf Goodness of embedding}\\
In the text we have already discussed the effective number of 
correctly embedded objects as a measure of `global goodness of
embedding.'   This measure, however, cannot tell us the embedding
quality  of each object.  It is often the case that the majority of
objects are embedded well even without sensitive dependence on
the initial conditions, but there are a few objects that 
consistently refuse to be embedded stably.  To judge the 
quality of embedding for each object $j$ we define 
\[
\Delta(j) \equiv \sum \left[T_{n(j)}(j) -n(j)\right]^2,  \label{D}
\]
Here, $n(j)$ is the rank order of $\d(i,j)$ among $N-1$ 
pairs $(i,j)$ for a given  $j$, and $T_{n(j)}(j)$ is the rank order of
$d(i,j)$ among $N-1$ pairs $(i,j)$ for the same $j$.  

$\Delta(j)$ can be regarded as 
a statistical variable
for the relative position of the $j$-th object with respect to the
remaining objects (Lehmann 1975).  We
can estimate the probability $P(\e)$ of $\Delta(j)<\e$ with the null 
hypothesis that the rank ordering of $d_{ij}$ ($ i \in \{1,2, \cdots,
N\}\setminus \{j\}$) 
is totally random with respect to the rank ordering of 
$\delta_{ij}$ ($ i \in \{1,2, \cdots, N\}\setminus \{j\}$). 
If $N$ is sufficiently large, then $\Delta(j)$ obeys the  normal
distribution 
with mean $(M^3-M)/6$ and variance $M^2(M+1)^2(M-1)/36$, where $M \equiv N-1$.  For
smaller $N$ there is a table for $P(\epsilon)$ 
(Lehmann 1975).  Thus, we can test the null hypothesis with a given
confidence level for $j$-th
object.

\noi
{\bf APPENDIX II}\\
{\bf Limitations and capabilities of linear methods}\\
\noi
The limitations and capabilities of PCA with
and without data preprocessing are illustrated
in this appendix.  There is no fundamental difference between PCA and SVD.
We consider the following artificial data $\{s_{gt}\}$, where 
$g$ ($=1,\cdots,517$) denote genes and $t$ ($=1, \cdots, 11$) the observation times:\\
$ $[Data set 1]
\[
s^1_{gt} = C_g \cos (2 \pi t / 11 + 2 \pi \delta_g).
\]
$ $[Data set 2]
\[
s^2_{gt} = \exp(s^1_{gt}).
\]
$ $[Data set 3]
\begin{eqnarray*}
s^3_{gt} &= &C_{g1} \cos (2 \pi t / 11 + 2 \pi \delta_{1g}) \\
& & + C_{g2} \exp[\cos (2 \pi t / 11 + 2 \pi \delta_{2g})] \\
& & \mbox{}\mbox{} + C_{g3}/ \cos (2 \pi t / 11 + 2 \pi \delta_{3g}).
\end{eqnarray*}
In the above, 
$C_g, \delta_g, C_{ig},\delta_{ig},(i=1,2,3)$ are uniform random numbers in
$[0,1]$.  That is, Data set 1 is a set of sinusoidal waves with random amplitudes and phases, Data
set 2 is the nonlinearly distorted Data set 1, and Data set 3 is a set of periodic functions that are
very different from simple oscillatory behaviors.

These data sets are analyzed by the following methods.\\
\noi
Method 1: PCA with the preprocessing used by Holter {\it et al}. (2000).
The preprocessing procedure is as follows:\\
step 1: Subtract the average,
$$
s'_{gt}= s_{gt} - \langle s_{gt}\rangle_{g,t},
$$
where  $\langle \bullet \rangle_{g,t}$ is the average over all genes and
experiments,
$$
\langle \bullet \rangle_{g,t} \equiv  \frac{\sum_{g,t} \bullet}{\sum_{g,t}1}.
$$
\\
step 2: (Column normalization) Normalize the data as
$$
s''_{gt} = \frac{s'_{gt}}{\sqrt{\left[\sum_g (s'_{gt})^{2}\right]}} .
$$\\
step 3: (Row normalization) Normalize the data as
$$
s'''_{gt} = \frac{s''_{gt}}{\sqrt{\left[\sum_t ( s''_{gt})^{2}\right]}} .
$$\\
Repeat these steps until the following condition is satisfied,
$$
\sqrt{\langle \left[s_{gt}-s'''_{gt}\right]^2 \rangle_{g,t}} < 0.01.
$$
\\
From the resultant $s_{gt}$ correlation matrix 
$Matr.(Cor_{t t'})$ 
is constructed, and then PCA is performed.

\noi
Method 2: PCA with the preprocessing 
so that $\sum_t s_{gt} =0 $ and $\sum_t s_{gt}^2=1$ for all $g$.  Of course, no iteration is needed for 
this preprocessing.  From the resultant $s_{gt}$ correlation matrix
$Matr.(Cor_{t t'})$  is constructed, and then PCA is performed.

\noi
Method 3: nMDS as done in the text.  That is, the negative of the correlation 
coefficient $Cor_{g g'}$ is used as the dissimilarity and nMDS is applied 
straightforwardly. Needless to say, no preprocessing of data is needed.

\marginparsep -9mm
\marginpar{\fbox{Fig. 6}}
\marginparsep 4mm

The results are exhibited in Figure 6.  The conclusions may be:\\
(1) For Data set 1, any method will do.\\
(2) For Data set 2, the procedure recommended by Holter {\it et al.} (2000) fails, 
although ironically simpler Method 2 still works very well.  If the amplitude $C$ 
is distributed in $[0,5]$ instead of $[0,1]$ (that is, the extent of the nonlinear distortion is increased), 
Method 2 becomes inferior to Method 3, 
but still Method 2 is adequate.\\
(3) For Data set 3, even Method 2 fails. nMDS (Method 3) still exhibits a ring-like 
structure. The method recommended by Holter {\it et al}.\ (2000) is obviously out 
of question.

Thus, we may conclude that nMDS is a versatile and all around data 
mining method for analyzing periodic temporal data.  Furthermore, 
we can point out that the preprocessing method in Method 1 should 
not be used because it could severely distort the original data (as 
may have been expected from the figures).
Suppose there are $N$ genes and $4$ time points.  Consider the 
following example (for the counterexample sake).
The first gene
has $(a,b,-b,-a)$ ($a>b>0$), and the remaining genes are all give
by $(1, 0,0,-1)$.  The $N \times 4$ matrix made from these 
vectors is polished by an iterative row and column vector 
normalization procedure.  If $N$ is sufficiently large, the
first row converges to $(0,1,-1,0)$ and the rest to $(1,0,0,-1)$, 
independent of $a$ and $b$.  If $b$ is small, then all the vectors
should behave almost the same way, but after polishing the 
out-of-phase component in the discrepancy between the first row and
the rest is dramatically enhanced, resulting in a spurious out of
phase temporal behavior.  Although the preceding exercise is 
trivial, the result  warns us the danger of using the so-called
polishing.  

\noi
{\bf REFERENCES}\\
\noi
Borg, I.,  and Groenen, P.  (1997) {\it Modern Multidimensional
 Scaling}, Springer, New York.

\noi
 Cox, T. F., and Cox,  M. A. A., (1994) {\it Multidimensional Scaling},
Chapman \& Hall, London.

\noi
Donoho, D.L.,  Vetterli,M.,  DeVore,R.A.,  and Daubechies, I. (1998)
Data Compression and Harmonic Analysis, 
{\it IEEE Transactions on Information Theory}, {\bf 44}, 
2435-2476.

\noi
Eisen, M. B., Spellman, P.T., Brown, P. O., and Botstein D., (1998)
    Cluster analysis and display of genome-wide expression patterns 
{\it Proc. Natl. Acd. Sci. USA}, {\bf 95} 14863-14868.

\noi
Green, P.E.,  Carmone Jr., F. J.,   and Smith, S. M., (1970) 
{\it Multidimensional Scaling : Concepts and Applications},
Allyn and Bacon, Massachusetts.

\noi
Hollander, M., and Wolfe, D. A., (1999)
{\it Nonparametric Statistical Methods},
 John Wiley \& Sons, New York.

\noi Holter, N. S., Mitra, M., Maritan, A., Cieplak, M., Banavar,
J. R., and Fedoroff, N. V.,  (2000)
 Fundamental patterns underlying gene expression profiles: Simplicity from complexity
{\it Proc. Natl. Acad. Sci. USA},  {\bf 97}, 8409-8414.

\noi
Iyer,  V. R.,   Eisen, M. B.,   Ross,  D. T., 
 Schuler,  G.,  Moore, T., L.,   Jeffrey C. F.   Trent,  J. M.,
Staudt, L. M.,  Hudson Jr.,   J.,   Boguski,   M. S.,   
Lashkari, D., Shalon, D.,   Botstein,  D.,  and  Brown,    P. O.,  (1999) 
The Transcriptional Program in the Response of Human Fibroblasts to Serum,
{\it Science}, {\bf 283}, 83-87.

\noi
Johansson, D.,  Lindgren, P.,  and Beglund, A.,  (2003)
A multivariate approach applied to microarray data for 
identification of genes with cell cycle-coupled transcription, 
{\em Bioinformatics}, {\bf 19}, 467-473.

\noi
Kanaya, S., Kinouchi M., Abe, T., Kudo, Y., Yamada, Y., Nishi, T., Mori,
Mori, H. and Ikemura, T., (2001) Analysis of codon usage diversity of
bacterial genes with a self-organizing map (SOM): characterization of
horizontally transferred genes with emphasis on the E. coli O157 genome.
{\it Gene}, {\bf 276}, 89-99.

\noi
Kasturi, J., Acharya, R.,  and Ramanathan, M., (2003)
An information theoretic approach for analyzing temporal patterns of gene
expression, {\it Bioinformatics}, {\bf 19},  449-458.

\noi Kruskal, J. B., (1964a) Multidimensional scaling by optimizing goodness of
fit to a nonmetric hypothesis, {\it Psychometrika}, {\bf 29}, 1-27.

\noi Kruskal, J. B., (1964b) 
Nonmetric multidimensional scaling: A numerical method, {\it Psychometrika},
{\bf 29}, 115-129.

\noi
Lehmann, E. L., (1975) {\it Nonparametrics}, Holden-Day, San Francisco.

\noi Shepard, R. N., (1962a) 
The analysis proximities: Multidimensional scaling
with an unknown distance function, I {\it Psychometrika}, {\bf 27}, 125-140.

\noi Shepard, R. N., (1962b) 
The analysis proximities: Multidimensional scaling
with an unknown distance function, II {\it Psychometrika}, {\bf 27}, 219-246.

\noi
Slonim, D. K., (2002) From patterns to pathways: gene
expression data analysis of age, {\it Nature Genetics
Supplement}, {\bf 32}, 502-508.

\noi
Spellman, P. T.,  Sherlock, G., Zhang, M. Q.,
Iyer, V. R., Anders, K.,  Eisen, M. B.,
Brown, P. O., Botstein, D., and Futcher, B.,  
(1998)
 Comprehensive Identification of Cell Cycle regulated Genes of the Yeast
{\it Saccharomyces cerevisiae}
by Microarray Hybridization,
{\it Molecular Biology of the Cell}, {\bf 9},  3273-3297.

\noi
Taguchi, Y-h., and  Oono, Y., (1999), unpublished.
\\http://www.granular.com/MDS/src/paper.pdf

\noi
Taguchi, Y-h., Oono, Y., and Yokoyama, K., (2001)
New possibilities of non-metric multidimensional scaling,
{\it Proc. Inst. Stat. Math.}, {\bf 49}, 133-153 (in Japanese).

\pagebreak
\onecolumn

\noindent
Figure legends

\noindent
Figure 1:
PCA results using correlation coefficient matrix. 
The first two principal components are used as the horizontal and
vertical axis, respectively (the cumulative proportion is 70 \%).
Genes whose experimental values are larger than $3.2$ are drawn
using filled boxes, otherwise drawn using small dots
(the corresponding color figure is available online).
From the top the time is, respectively, 15 min, 30 min, 1 hr, 2 hr,
4 hr, 6 hr, 8 hr, 12 hr, 16 hr, 20 hr, 24 hr.

\noindent
Figure 2.
Two dimensional embedding  result obtained by nMDS.

\noindent
Figure 3:
(a) Temporal patterns of gene expression levels visualized with the aid of
nMDS.
Colors indicate relative intensity of experimental values
normalized so that $\sum_t s_{gt}=0$
and $\sum_t s_{gt}^2 =1$ where $s_{gt}$ is experimental variable
of $g$th genes at time $t$. 
(red $> 1.6$, yellow $> 1.2$, green $> 0.8$, pale blue $> 0.4$, 
gray $< 0.4$).
Time sequences are the same as explained at Fig. 1.
\noindent
(b) Gene expression data as a function of the angle measured from the 
vertical axis in (a).
The 
horizontal axis corresponds to $t$. The color convention is the same as in (a).

\noindent
Figure 4
Comparison between the nMDS embedding results with 
two different initial configurations after Procrustean similarity
transformation.
The horizontal (resp., vertical) coordinates 
are compared in the  left (resp., right) figure.
In each figure $x$-axis corresponds to the result from 
one initial condition and the $y$-axis the other.

\noindent
Figure 5:
3D unfolding of the temporal pattern of gene expression level with the aid of
nMDS (3D).
Experimental values are normalized as explained in Fig.\ 3.
Genes whose experimental values are larger than $1.6$ are drawn
using filled boxes, otherwise drawn using small dots
(the corresponding color figure is available online).
The horizontal (resp., vertical) axis represents $\phi$ (resp., $\theta$).
See the text for detail.

\noindent
Figure 6 Comparison of linear and nonlinear methods.\\
Method 1: PCA with polishing (Holter {\it et al}.\ 2000);
Method 2: PCA with normalization;
Method 3: 2$D$ space embedding with the aid of nMDS.
See the text for Data sets and Methods.
For Methods 1 and 2, horizontal and vertical axes
are the first and second principal components, respectively, and
the percentages describe cumulative proportions.
For Method 3, the percentages are the indicators of goodness defined in the
text.

\pagebreak
\noindent
Figure legends for online only color figures:

\noindent
Figure 1 (Color; online only):
PCA results using correlation coefficient matrix. 
The first two principal components are used as the horizontal and
vertical axis, respectively (the cumulative proportion is 70 \%).
Colors indicate relative intensity of experimental values
(red $> 3.2$, yellow $> 2.4$, green $> 1.6$, pale blue $> 0.8$, 
gray $< 0.8$). 
From the top the time is, respectively, 15 min., 30 min., 1 hr, 2 hr,
4 hr, 6 hr, 8 hr, 12 hr, 16 hr, 20 hr, 24 hr.

\noindent
Figure 5 (Color: online only):
3D unfolding of the temporal pattern of gene expression level with the aid of
nMDS (3D).
The color convention is the same as explained in Figure 3.
The horizontal (resp., vertical) axis represents $\phi$ (resp., $\theta$).
See the text for detail.

\end{document}